\documentclass[a4paper,12pt]{article}
\usepackage[utf8]{inputenc} 
\usepackage[english]{babel} 
\usepackage{amsfonts}
\usepackage[top=3cm,left=3cm,right=2cm,bottom=2cm]{geometry}
\usepackage{graphicx}
\usepackage{icomma}
\usepackage{array}
\newcommand{\Rho}{P}
\newcommand{\adot}{\dot{a}}
\newcommand{\phidot}{\dot{\phi}}
\newcommand{\phiddot}{\ddot{\phi}}

\begin{document}

\title{\large\bf Inflationary Cosmology in Scalar-Tensor Gravity: Reconstructing Higgs-Like Potentials}
\author{Anderson M. Silva${}^{*}$ \& Jim E. F. Skea${}^{\dag}$}

\maketitle


\begin{center}
$^{*}$Programa de Pós-graduação em Física, Instituto de Física Armando Dias Tavares, Universidade do Estado do Rio de Janeiro, Rua São Francisco Xavier, 524, Maracanã, Rio de Janeiro -- RJ.

$^{\dag}$ Departamento de Física Teórica, Instituto de Física Armando Dias Tavares, Universidade do Estado do Rio de Janeiro, Rua São Francisco Xavier, 524, Maracanã, Rio de Janeiro -- RJ.

\end{center}

\centerline{emails: ${}^{*}$ams.jhss@gmail.com, ${}^{\dag}$jimskea@gmail.com}

\abstract{
We study cosmology in scalar-tensor (Bergmann Wagoner) gravity, restricting the coupling function, $\omega(\phi)$
to be constant. Rather than specify the form of the cosmological function, $\lambda(\phi)$, the scalar field is modelled as a function which decays
to its present value $\phi=1$. Solutions of the field equations are found for which $\lambda(\phi)$ evolves from a large value (approximately 1)
near the singularity to a small,
non-zero value at later times, avoiding the problem of pre-inflationary collapse in standard general relativistic cosmology. Interpreting the
model within the framework of Brans-Dicke theory with a scalar potential, we find that for suitable initial conditions,
 the reconstructed potential at late times
for flat, open and closed universes is well described by a Higgs-like Mexican hat potential, quartic in $\phi$, though this was not built in to the initial assumptions.
}


\section{Introduction}

For some time now it has been recognised that ``old inflation''~\cite{Guth}, in which inflation is generated by a scalar field,
suffers from the shortcoming that inflation begins some time distant from the initial singularity, restricting the initial conditions of the
Universe to those for which collapse does not occur before the phase transition that generates inflation.
However Linde's chaotic inflation~\cite{Linde} and ``new inflation'' models avoid this difficulty by producing inflation soon after the Big Bang.
Many new inflation models assume that the potential for the scalar field has a particular form (see~\cite{LindeReview} for a recent review). 

In this paper, we study cosmology in scalar-tensor gravity, also called Bergmann-Wagoner (BW) theory~\cite{Bergmann,Wagoner}
which naturally contains a variable cosmological function, $\lambda(\phi)$ - or running cosmological `constant' -
in addition to a coupling parameter, $\omega(\phi)$, which may also depend on the scalar field. In this theory, as in
Brans Dicke theory, the scalar field is associated with the gravitational coupling such that $\phi=G^{-1}$.

We  search for solutions that have various desirable properties:
{\bf (1)}
``natural'' initial conditions, in the sense that values of physical
quantities near the singularity are of order 1 in Planck units; {\bf (2)}
sizable inflation is produced directly after the Big Bang; {\bf (3)}
they evolve to a final state with a small, but non-zero, cosmological ``constant'';
{\bf (4)} the value of $\omega$ is  compatible with observations;
{\bf (5)} they produce a variation in  $G$ at late times which is compatible with present-day observations.

To this end we choose a model for $\phi(t)$, rather than a model for $\lambda(\phi)$, simply demanding that
it evolve from some initial almost constant value to a late-time almost constant value (which will be unity in
Planck units). We find that a relatively small change in $\phi$, of some $10\%$  is sufficient to produce solutions
with the above properties. When we reconstruct the dependence of $\lambda(\phi)$ and the equivalent
potential $U(\phi)$, we find that, at low energies, the latter is very well described by a
Mexican hat/Higgs potential, with the scalar field evolving
towards a local maximum at $\phi=1$.

\section{Scalar Tensor Gravity} 

The scalar-tensor field equations for BW theory are derived from the variational principle applied to the action~\cite{Will} 
\begin{equation}
\begin{array}{l}
\displaystyle
I_{\rm BW} = \frac{1}{16\pi}\int \left [\phi R -\frac{\omega(\phi) \phi_{,\mu}\phi^{,\mu}}{\phi}
+2\phi\lambda(\phi)\right ]\times
\\
\displaystyle
\sqrt{-g}\,{\rm d}^4 x+ I_{\rm NG}(q_A, g_{\mu\nu})
\end{array}
\label{BWAction}
\end{equation}
where $\omega(\phi)$ is the coupling parameter,  $\lambda(\phi)$ the cosmological function
and $I_{\rm NG}$ describes the non-gravitational part of the action (throughout the paper we use reduced Planck units with $c= G=\hbar=1)$.
As with Brans-Dicke (BD) theory, the gravitational constant of General Relativity is replaced by a function $G(t,\vec{r})=\phi^{-1}$,  
where $\phi$ is a scalar field, with the  BD cosmological constant and coupling constant generalized to functions $\lambda(\phi)$ and $\omega(\phi)$ respectively.
Using the subscript $t$ to denote present-day values, and assuming homogeneity, in Planck units $G_t=1\Rightarrow\phi_t=1$.
Observational evidence~\cite{Turyshev2004,Kaspi,RothmanMatzner} suggests that today  $|\dot{G}/G|=|\dot{\phi}/{\phi}|<10^{-13}\, {\rm yr}^{-1}$,
or in Planck units  $|\dot{\phi}/{\phi}|<10^{-64}\,t_{\rm Pl}^{-1}$.

BW theory generalizes BD theory, for which the gravitational part of the action may be written
\begin{equation}
	I_{\rm BD} = \frac{1}{16\pi}\int \left [\phi R - \omega_{\rm BD} \frac{\phi_{,\mu}\phi^{,\mu}}{\phi}\right]\sqrt{-g}\,{\rm d}^4 x
\label{BDAction}
\end{equation}
where the coupling parameter $\omega_{\rm BD}$ is now constant, and $\lambda(\phi)=0$.

If a scalar potential $U(\phi)$ is added to the BD action then the gravitational action
 becomes~\cite{LeeKimMyung}
\begin{equation}
\begin{array}{l}
\displaystyle
I_{\rm BD} = \frac{1}{16\pi}\int
\left [\phi R - \omega_{\rm BD} \frac{\phi_{,\mu}\phi^{,\mu}}{\phi}+16\pi U_{\rm BD}(\phi)\right]
\times\\
\displaystyle
\sqrt{-g}\,{\rm d}^4 x
\end{array}
\label{BDAction2}
\end{equation}
and we see that BW theory with constant $\omega$ is equivalent to BD theory plus a scalar potential with the identification
\begin{equation}
U_{\rm BD}(\phi) = \phi\lambda(\phi) / 8\pi.
\label{Uphi}
\end{equation}

Though the theories are equivalent, we prefer the BW approach with the interpretation of a time-varying cosmological
function (or running cosmological constant)
as it lends itself more clearly to our analysis, as well as leaving open the possibility of studying the effect of a variable $\omega$. 

The field equations  derived from~(\ref{BWAction}) are
\begin{equation}
\begin{array}{l}
\displaystyle
G_{\mu\nu}-\lambda(\phi)g_{\mu\nu}= \frac{8\pi}{\phi}T_{\mu\nu}+
\frac{\omega(\phi)}{\phi^2}\phi_{,\mu}\phi_{,\nu}\\
\displaystyle
-\frac{1}{2}g_{\mu\nu}\phi_{,\lambda}\phi^{, \lambda}
+ \frac{1}{\phi}\,(\phi_{;\mu\nu} -g_{\mu\nu}\Box\phi),
\end{array}
\label{field}
\end{equation}
and
\begin{equation}
[3+2\omega(\phi)]\Box\phi+2\phi^2\frac{{\rm d}\lambda}{{\rm d}\phi} -2\phi\lambda(\phi)
=8\pi T - \frac{{\rm d}\omega}{{\rm d}\phi}\phi_{,\mu}\phi^{,\mu}.
\label{ephi}
\end{equation}
We note that, different from pure BD theory, where $\omega=-3/2$ implies $T=0$,
effectively demanding a vacuum or radiation fluid, there is no such restriction in BW
theory if $\lambda\neq 0$.

The standard approach in scalar-tensor cosmology is to specify the potential $U(\phi)$, or equivalently $\lambda(\phi)$, and study the subsequent evolution
of the variables in the model. Here we adopt a different procedure: we model the evolution of $\phi(t)$ as a function that decays from
an initial value, $\phi(0)=\phi_0$ to its present-day value $\phi_t=1$, hoping to  identify solutions that generate inflation from $t = 0$,
with $\lambda(t)$ evolving to a smaller, constant value as $t\rightarrow\infty$. Once these solutions are identified we reconstruct the
functions $\lambda(\phi)$ and $U(\phi)$.  

We assume that the energy-momentum tensor is that of a perfect fluid
\begin{equation}
T_{\mu\nu}=(\rho+p)u_\mu u_\nu -pg_{\mu\nu},
\end{equation}
with $\rho$ the energy density, $p$  the pressure of matter and $u$ the four-velocity with $u^\mu u_\mu =1$.

Supposing a  homogeneous and isotropic universe, all variables depend only on time, $t$, and
the space-time is described by the Friedmann Lemaître Robertson Walker (FLRW) metric in the form
\begin{equation}
ds^2=dt^2-a^ 2(t)\left[\frac{dr^ 2}{1-kr^ 2}+r^ 2(d\theta^2+\sin^ 2\theta d\varphi^ 2)\right].
\end{equation}

Since we are mainly interested in evolution close to the Big Bang, we use the equation of state $p=\rho/3$.
To reduce the amount of freedom in the model, we impose $\omega(\phi)$ constant (hence making the model
equivalent to BD plus scalar potential). Using an overdot to
denote a time derivative
the field equations for $G_{00}$ and $G_{11}$ are, respectively,
\begin{equation}
\frac{\dot{a}^2}{a^2}+\frac{k}{a^2}-\frac{\lambda(\phi)}{3}=\frac{8\pi\rho}{3\phi}+\frac{\omega}{6}\frac{\dot{\phi}^2}{\phi^2}-\frac{\dot{a}}{a}\frac{\dot{\phi}}{\phi},
\label{eq1}
\end{equation}
\begin{equation}
2\frac{\ddot{a}}{a}+\frac{\dot{a}^2}{a^2}+\frac{k}{a^2}-\lambda(\phi)=-\frac{8\pi \rho}{3\phi}-\frac{\omega}{2}\frac{\dot{\phi}^2}{\phi^2}-2\frac{\dot{a}}{a}\frac{\dot{\phi}}{\phi}-\frac{\ddot{\phi}}{\phi}.
\label{eq2}
\end{equation}

Considering  $\lambda(\phi(t))$ as $\lambda(t)$ equation~(\ref{ephi}) may be rewritten
\begin{equation}
(3+2\omega)\left(\dot{\phi}\ddot{\phi} +3\dot{\phi}^2\frac{\dot{a}}{a}\right)+2\phi^2\dot{\lambda} -2\phi\dot{\phi}\lambda=0 
\label{eq3}
\end{equation}

Finally, the Bianchi identity $\nabla_\mu G^{\mu\nu}=0$ yields 
\begin{equation}
\begin{array}{l}
\displaystyle
8\pi\left[\frac{{\rm d}}{{\rm d}t}\!\left(\frac{\rho}{\phi}\right)+4\frac{\adot}{a}\,\frac{\dot{\phi}}{\phi}\right]
+\frac{\rm{d}}{\rm{d} t}\!\left(\lambda+\frac{\omega}{2}\frac{\dot{\phi}^2}{\phi^2}
-3\frac{\adot}{a}\,\frac{\phidot}{\phi}\right)\\
\displaystyle
+3\frac{\adot}{a}\left(\omega\,\frac{\phidot^2}{\phi^2}-\frac{\adot}{a}\,\frac{\phidot}{\phi}
+\frac{\phiddot}{\phi}\right)=0
\label{eq4}
\end{array}
\end{equation}
Defining $H(t)=\dot{a}/a$, $\Phi(t)=\dot{\phi}/\phi$ and $\Rho(t)= 8\pi\rho/\phi$,
subtracting~(\ref{eq1}) from~(\ref{eq2}) and rearranging  we have
\begin{equation}
\dot{H}=-H^2+\frac{1}{3}(\lambda -\omega\Phi^2-\Rho)-\frac{1}{2}(H\Phi+\dot{\Phi}+\Phi^2),
\label{e1}
\end{equation}
\begin{equation}
\dot{\lambda}=\lambda\Phi - \left(\omega+\frac{3}{2}\right)(3 H\Phi^2+\Phi^3+\Phi\dot{\Phi}),
\label{e2}
\end{equation}
\begin{equation}
\dot{\Rho}=-\Rho(4H+\Phi).
\label{e3}
\end{equation}

Written in terms of $\rho$ this final equation is just the usual energy-momentum conservation equation with solution $\rho=\rho_0 a^{-4}$.

We are searching for solutions where $\lambda(\phi)$ decays from a large initial value $\approx 1$ to a much smaller value, with two-tier
inflation. This suggests a model in which $\phi$ evolves from an initial value $\phi_0$ near the
singularity (which we take as $t=0$ rather than $t=1$)
to its present value of $\phi=1$, with  $\phi$ approximately constant for long periods around these values.
Between these two eras, the intermediate decay of $\phi$, occurring around $t=t_d$ say,
could be caused by some phase transition at the corresponding energy.

A model for $\phi(t)$ with such a smooth behaviour is
\begin{equation}
\phi(t)=\frac{1}{2}(1+\phi_0)+\frac{1}{2}(1-\phi_0)\tanh[c(t-t_d)],
\label{phi}
\end{equation}
where $c$ governs the rate of decay.
For this function
 $\phi$ (and consequently $\lambda$) is approximately constant around $t=0$, decaying smoothly
around $t_d$, and tending asymptotically to $\phi=1$.
 Since $\phi$ tends to a constant value for $t\rightarrow\infty$, $\lambda(\phi)$ will also tend to a constant, in general
different from $\lambda(0)$.

This cosmology thus depends on the four parameters $\omega$, $c$, $\phi_0$ and $t_d$, as well as the values of $\lambda$, $\Rho$ and $H$ at $t=0$.
Siince $\omega$ is assumed constant, we can use present-day limits, the most stringent of which are
 given by the analysis of signals from the Cassini spacecraft~\cite{Bertotti}. These limit the PPN parameter
$\gamma = 1+ (2.1\pm 2.3) \times 10^{-5}$. From the relation~\cite{Will}
\[
\gamma = \frac{1+\omega}{2+\omega}
\]
we have that, to two standard errors, $\omega < -15,000$ or $\omega  > 40,000$.

Some articles on scalar-tensor cosmology~\cite{LeeKimMyung,MakHarko} use values of $\omega$ which are negative
with small modulus (typically around $\omega=-3/2$), incompatible with the Cassini
observations. However fitting of type~I Supernova data~\cite{Fabris} suggests that  Brans-Dicke theory
with $\omega\approx -3/2$ describes these observations on cosmological scales slightly better than
the $\Lambda$CDM models.
This may be taken as evidence for a time-varying or spatially dependent $\omega$,
both of which fall within the scope of BW theory. Here, however, we maintain
$\omega$ constant and positive.

We integrate the equations from the Planck time with
all initial values close to unity.
Our aim is to determine whether there exist initial conditions which, 
for the proposed
model for $\phi(t)$, produce a cosmology with inflation from $t=0$, while
$\lambda$ decays to a smaller value compatible with an observed value of the cosmological ``constant''
today. Evidently it is unrealistic to integrate the $10^{60}$ Planck times until the present day, but we can study what
occurs for a reasonable number of Planck times around the period during which $\phi$ decays, for models with $k =0 ,\pm 1$.

\section{Results}

For the flat, open and closed cases we model $\phi$ by~(\ref{phi}) with the same values of $\phi_0 = 1.1$, $t_d = 700$ and $c=0.0074$.
This latter value produces a decay period (calculated as 95\% of the fall) that lasts approximately $500\, t_{\rm Pl}$. The time
in the middle of the decay, $t_d$,  is
chosen so that $\phi\approx\phi_0$ for a substantial time before it decays. We choose $\omega=50,000$ in all cases,
compatible with the limits stated in the previous section. The equations were integrated using  an adaptive
Runge-Kutta 8th/9th order method
due to Verner~\cite{Verner} implemented by the authors using
the multiple precision MPFR library~\cite{MPFR} with 1000 bits (300 decimal digits),
and an error per step of $10^{-30}$.

\begin{figure}[!htb]
\begin{center}
\includegraphics[width=0.4\textwidth]{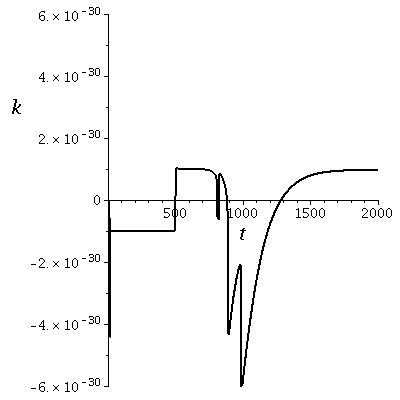}
\end{center}
\caption{Absolute error in $k$ for the flat model.}
\label{figk}
\end{figure}

To check that the numerical integrator was producing the desired precision, we evaluated
$k$ using~(\ref{e1}). For example the value of $|k|$ for the flat model studied below is shown in~Figure~\ref{figk},
confirming that the error is around $10^{-30}$, in line with expectations.
 
\subsection{Flat Universe}

\begin{figure}[!htb]
\begin{center}
\includegraphics[width=0.24\textwidth]{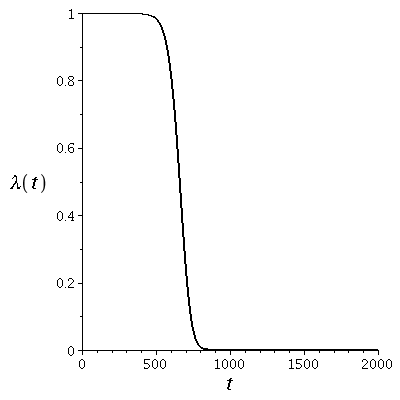}
\includegraphics[width=0.24\textwidth]{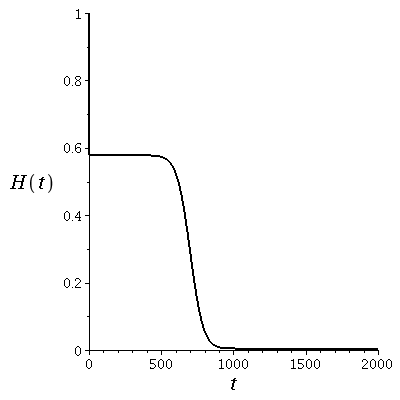}
\includegraphics[width=0.24\textwidth]{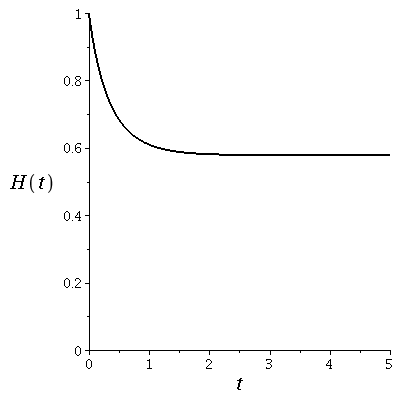} 
\includegraphics[width=0.24\textwidth]{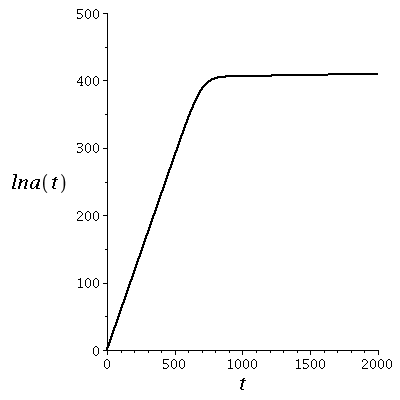}
\end{center}
\caption{Plots of $\lambda(t)$ and  $H(t)$ up to $t=2,000$, $H(t)$ near the initial singularity and $\ln a(t)$  
for the $k=0$ model.}
\label{figflat1}
\end{figure}

For $k=0$  the initial conditions were: $H=1$, $\Rho=2$, $a=1$ and  via~(\ref{eq1}),
we obtain $\lambda\approx 0.999999872$. In figure~\ref{figflat1} we show the evolution of $\lambda$, $H$,  and $\ln(a)$ 
from $t=0$ until $t=2,000$. A detailed view of $H(t)$ near $t=0$ is also given.

The first graph shows the evolution of $\lambda(t)$ from $t=0$ until $t=2,000$. We see that, during this period, $\lambda$
decays from approximately unity to $\approx 3.73\times 10^{-5}$. Integration up to $t=2\times 10^7$ does not significantly
change this value, which is evidently much higher than today's observed value of $10^{-122}$, but it does show that in a
short time $\lambda$ can decay significantly from a value of 1 near the singularity to a much smaller value.
As we shall see shortly,
other values of the parameters in the potential and initial conditions, the decay in $\lambda$ can be greater.

The second graph shows the variation of the Hubble parameter during the same period. We notice that $H(t)$ also decays,
but in two distinct steps, falling from 1 to 0.6 in just one Planck time (seen in detail in the third plot),
followed by a period when $H(t)$ is approximately
constant (a first inflationary period) up to $t\approx 600$. After this, $H$ decays abruptly (together with $\phi$),
reaching $H\approx 0.003541$ around $t=1,100$ and changing little after that: at $t=2\times 10^7$, $H\approx 0.003526$.
In this final stage, the Universe can be described as passing through a second, slower inflationary phase. During both
inflationary phases $H\approx\sqrt{\lambda/3}$ and we have (approximate) de-Sitter solutions.
With these initial conditions, the scale factor, $a(t)$, undergoes 400 e-folds
during the first inflationary phase, thereafter increasing at a much slower rate.

Hence, albeit on a different time scale,
we have the main characteristics needed to describe a Universe
in which an initial inflationary period helps solve the horizon and flatness problems, while at late times a
residual cosmological `constant' produces an accelerated
expansion.
Inflation
appears very soon after the universe emerges from the quantum era, thereby problems of early recollapse
are avoided;  and while, in General Relativity the primordial inflaton
and late-time cosmological constant are normally considered separate entities,  in BW theory they are 
naturally connected within the model.

\begin{figure}[!htb]
\begin{center}
\includegraphics[width=0.3\textwidth]{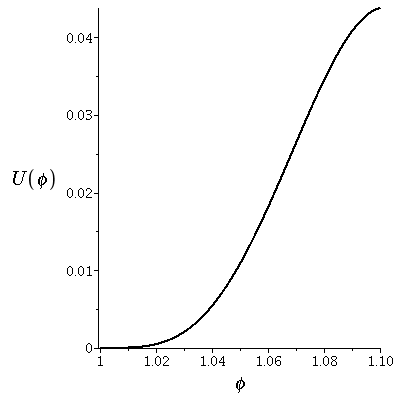}
\includegraphics[width=0.3\textwidth]{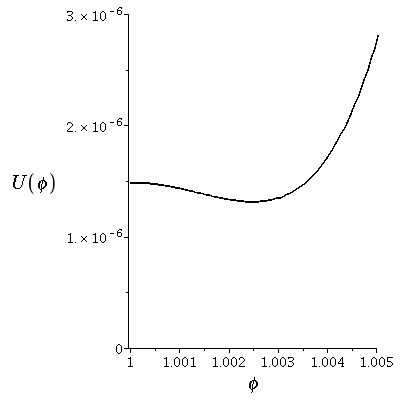}
\end{center}
\caption{The reconstructed potential $U(\phi)$: on the left a graph of the whole potential;
on the right a detailed view showing the minimum near $\phi=1$. The scalar field evolves from right to left.
}
\label{figU0}
\end{figure}

It is interesting to ask what type of potential in BD theory would produce this type of solution.
In figure~\ref{figU0} we reconstruct from $\phi$ and $\lambda$ the equivalent BD potential~(\ref{Uphi}).
The graph on the right shows in detail the low-energy behaviour of the potential,
which has a minimum away from $\phi=1$. This low-energy potential $\phi<1.005$ is
exceptionally well described by a one-dimensional Mexican Hat function around $\phi=1$ of the form
\begin{equation}
U(\phi) = A(\phi-1)^4 +B(\phi-1)^2 + C,
\label{Ufit}
\end{equation}
with $A=4377\pm 4$, $B=-0.5268\pm0.0001$ and $C=1.4832\times 10^{-6}\pm 2\times 10^{-10}$. The fit was performed
by Maple with 204 data points in the interval $\phi\in(1.0, 1.005)$. We note that the ratio $|B|/A\approx 10^{-5}$ while, for
the Higgs potential this ratio is $246,\,{\rm GeV}$ or $10^{-16}$. Different choices of the parameters 
in the model for $\phi$ produce potentials with different values of $A$, $B$ and $C$.
For example, since $U(\phi)=\phi\lambda$ and $\phi\approx 1$ at late times, a small late-time value of $\lambda$ is equivalent
to a small value of $U(1)=C$.

\begin{figure}[!htb]
\begin{center}
\includegraphics[width=0.24\textwidth]{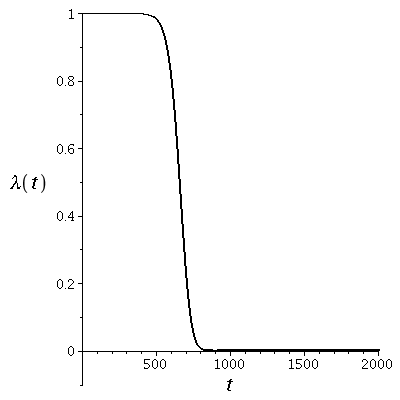}
\includegraphics[width=0.24\textwidth]{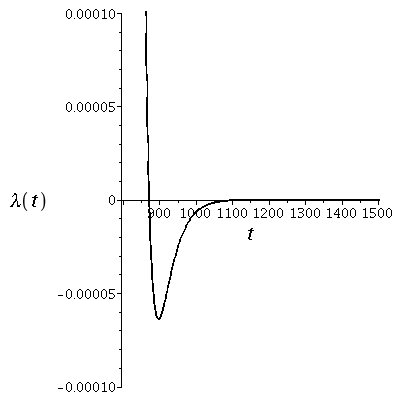}
\includegraphics[width=0.24\textwidth]{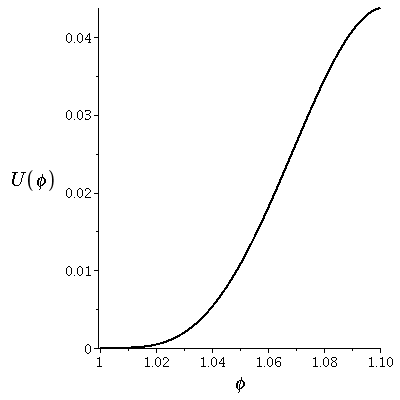}
\includegraphics[width=0.24\textwidth]{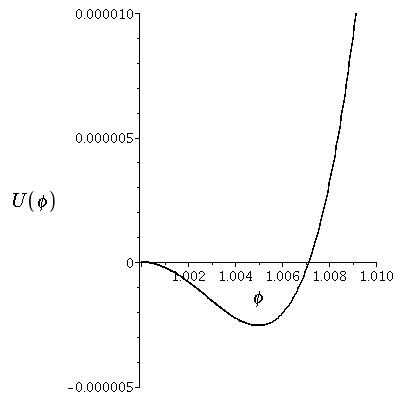}
\end{center}
\caption{The first two graphs show the evolution of $\lambda(t)$, with a detail of the period during which $\lambda < 0$; the second
pair of graphs shows the reconstructed potential, with a detail of the Mexican hat  behaviour near $\phi=1$}
\label{figflat2}
\end{figure}

It is not difficult to obtain smaller values of $\lambda$ at late times.
For example, by slightly altering the decay rate parameter in the potential to 
$c=7.44770532\times 10^{-3}$,
(for which $\lambda(0)\approx 0.9999998796$) we find that $\lambda_t$ falls by a factor of $10^{13}$
to $6.43\times 10^{-18}$, as plotted in the first two graphs of figure~\ref{figflat2}.  
We not there is a period between $t=850$ and $t=1,100$ when $\lambda(\phi)$ is negative, but
this does not lead to a recollapse.
Once again, for low energies ($\phi < 1.005$) the potential has the form~(\ref{Ufit})
with $A=4377$, $B=-0.2108$ and $C=2.56\times 10^{-19}$. It is interesting that
$\phi$ evolves towards a local maximum of the potential which is, in principle, an unstable equilibrium
point. However $\phi(t)$ tends to this maximum only as $t\rightarrow\infty$ so this instability is
effectively irrelevant.  

\subsection{\bf Closed Universe}

\begin{figure}[!htb]
\begin{center}
\includegraphics[width=0.24\textwidth]{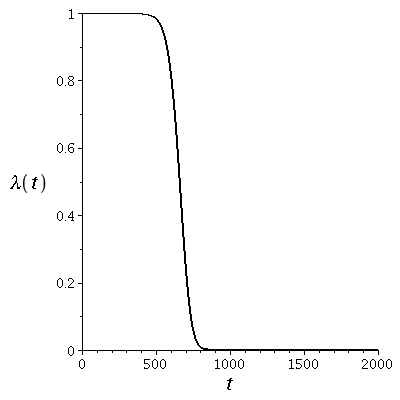}
\includegraphics[width=0.24\textwidth]{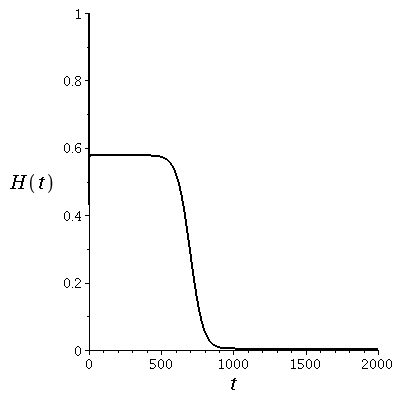}
\includegraphics[width=0.24\textwidth]{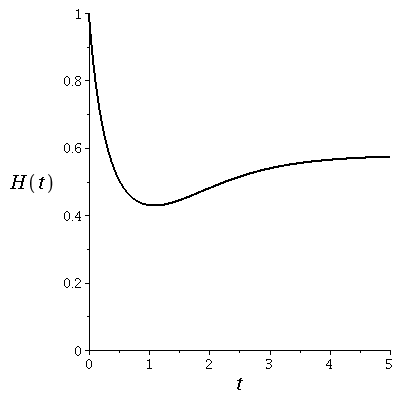} 
\includegraphics[width=0.24\textwidth]{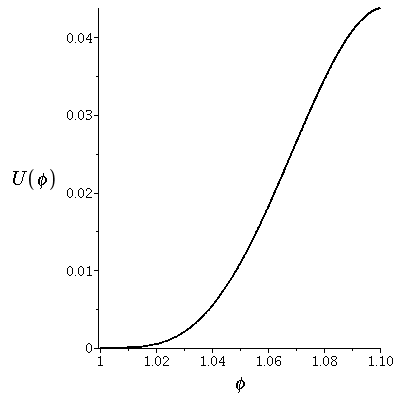}
\end{center}
\caption{Variation of $\lambda(t)$, the Hubble parameter $H(t)$,
a detail of $H(t)$ near $t=0$,    
and the potential $U(\phi)$ for the closed universe.}
\label{figkpositive}
\end{figure}

For $k=1$ the same parameters were used in the model of $\phi(t)$ with the initial conditions slightly modified
to satisfy the field equations with $k=1$, giving $\Rho(0)=5$ for the same value of $\lambda(0)$.
Figure~\ref{figkpositive} shows the results of the integration. The only significant difference in the results
is that the initial fall of $H(t)$ reaches a minimum around $t=1$ before recovering to a value of around 0.6.
The potential $U(\phi)$ is once again well described by the model~(\ref{Ufit}) with the same values of
the parameters. The evolution of $a(t)$ (not shown) again undergoes 400 e-folds
during the primordial inflationary phase.

\subsection{Open Universe}

\begin{figure}[!htb]
\begin{center}
\includegraphics[width=0.24\textwidth]{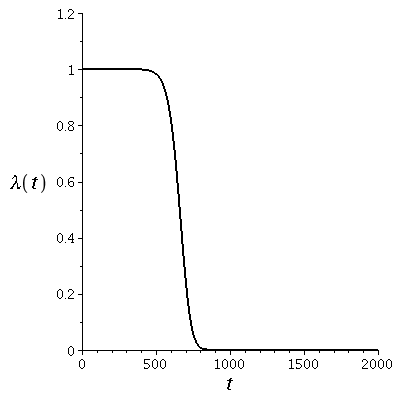}
\includegraphics[width=0.24\textwidth]{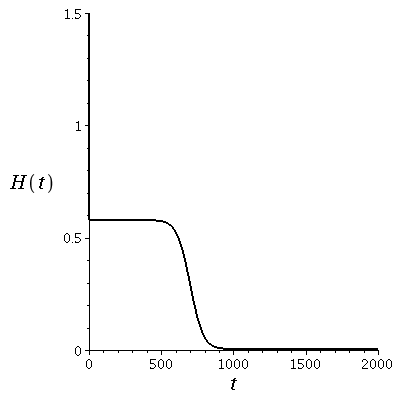}
\includegraphics[width=0.24\textwidth]{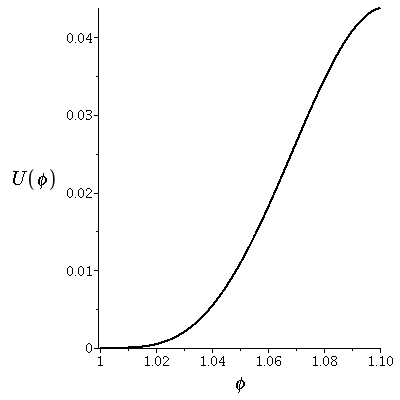}
\includegraphics[width=0.24\textwidth]{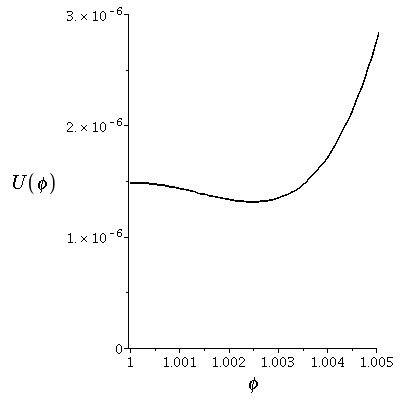}

\end{center}
\caption{Variation of $\lambda(t)$, the Hubble parameter $H(t)$, the potential $U(\phi)$ and a detail
of the low energy potential  for an open universe.}
\label{figopen}
\end{figure}

Exactly the same model for $\phi(t)$ was used in the analysis for $k=-1$, but with
different initial conditions: $H=1.5$, $\Rho=2.75$, $a=1$ and consequently
$\lambda(0)\approx 0.9999998194$. In
figure~\ref{figopen} we see that the results are initially very similar to the
flat model, but there is a larger final value of $\lambda \approx 3.7\times10^{-5}$, and consequently
larger values of $H(t)$ and a faster late-time expansion of $a(t)$, though again the variation
of $a(t)$ from the Big Bang until the end of the decay in $\phi$ is very similar to the $k=0$ cases.

\section{Conclusions}

The intention of this work is to study cosmological models within a restricted version
of the Bergmann-Wagoner theory of gravity for which the coupling parameter
$\omega(\phi)$ is constant. This restriction makes the theory
equivalent to Brans-Dicke theory with a scalar potential. A simple model
for the scalar field was proposed in which $\phi(t)$ is approximately constant
in the early and late stages, passing smoothly through a period when it rapidly
decays. For open, closed and flat FLRW models, solutions were sought and found in
which the cosmological function, $\lambda(\phi)$ evolves from a value close to unity
(in Planck units) near the Big Bang to a smaller almost constant value at late times.
Because of computational restrictions, the evolution was studied over `short' time scales
(in cosmological terms) but we believe that this behaviour can be reproduced over longer
time scales by appropriate choices of the parameters and initial conditions.

It was hoped that the problem of the fine tuning of initial conditions which beset
 standard inflation would be solved with a natural initial value of $\lambda\approx 1$
driving primordial inflation to avoid an early collapse. In the desired scenario, $\lambda$
would subsequently decay naturally to a value close to, but not exactly, zero, compatible with
present-day observations. On the positive side, solutions with these characteristics have been
determined. However
different choices of initial conditions and/or parameters in the scalar field model can produce
other sorts of behaviour, such as early collapse.

Though our approach was initially couched in terms of a running cosmological function
$\lambda(\phi)$,
within Bergmann-Wagoner theory, an alternative point of view in which $\lambda$  is
substituted by an equivalent potential $U(\phi)$ within Brans-Dicke theory was also
investigated. Curiously, it was found that the equivalent potential at low energies
has a form of a Higgs-like or Mexican hat potential, centred around $\phi=1$, with
the BD scalar field evolving towards the local maximum there. In some sense this
is to be expected from our model of $\phi(t)$: since we demand that $\phi$
tends towards $\phi=1$ as $t\rightarrow\infty$, the late-time potential must be
 $U(1)$. What is unexpected, and certainly not built into the assumptions,
is the form of $U(\phi)$ at low energies and the fact that $\phi=1$ corresponds
to a local maximum of $U$. It is interesting to speculate whether or not other
forms of $\phi(t)$ with similar early and late-time behaviours are associated with
similar potentials.

\end{document}